\documentstyle[prl,aps]{revtex}

\begin{document}
\draft
\title{Comment on the Erratum: Multiparticle Bose-Einstein Correlations\\
	$\left[\right.$ Phys. Rev. C 61, 029902(E) (2000) of
	Phys. Rev. C 57, 3324 (1998) $\left. \right]$
	}

\author{T. Cs\"org\H o$^{1,2}$
	\thanks{E-mails: csorgo@sunserv.kfki.hu,
	csorgo@nt1.phys.columbia.edu}{\ }
}
\address{
{$^1$Department of Physics, Columbia University,
 538 W 120-th Street, New York, NY 10027 }\\
{$^2$MTA KFKI RMKI, H-1525 Budapest 114. POB. 49, Hungary}
 }
\date{\today}
\maketitle

Although the Erratum~\cite{urs-err} correctly
states that the solution of model (4.1) depends only on two
effective parameters, in contrast to earlier statements of
ref.~\cite{urs-mpbe},
the Erratum presents the formulas for
$R_{{\rm eff}}$ and $\Delta_{{\rm eff}}$ incorrectly,
as can be seen even from a dimensional analysis.
The correct formulas for model (4.1)
were given in eq. (16) of ref. ~\cite{cstjz},
using a notation $R_{{\rm eff}} = R_e$ and
$\Delta_{{\rm eff}} = \sigma_T$,
as introduced in ref.~\cite{balazs}.
The analytic solution of model (4.1) was summarized in ref.~\cite{cstjz},
detailed investigations of the model properties
were described in refs.~\cite{jzcst,plcoh,cs-nato}.

Supported by the US - Hungarian Joint Fund MAKA 652/1998,
by the OTKA grants T025435 and T029158, and by the
US Department of Energy Contracts No.
DE - FG02 -93ER40764, DE-FG-02-92-ER40699 and
DE - AC02 -76CH00016.

\end{document}